\documentclass[
aps,%
12pt,%
final,%
notitlepage,%
oneside,%
onecolumn,%
nobibnotes,%
nofootinbib,%
superscriptaddress,%
noshowpacs]%
{revtex4}%
\usepackage{graphics}
\usepackage[english]{babel}
\usepackage{amssymb,amsmath}

\begin{document}
\newcommand{\rmbf}[1]{{\rm\bf #1}}

\title{BC\_NPI module for the analysis \\ of $B_c \to J/\psi +n\pi$ and  $B_c \to B_s +n\pi$ decays \\ within the EvtGen package.}

\author{A.V. Berezhnoy}

\email{Alexander.Berezhnoy@cern.ch}

\affiliation{SINP of Moscow State University, Russia}

\author{A.K. Likhoded}

\email{Anatolii.Likhoded@ihep.ru}

\affiliation{Institute for High Energy Physics, Protvino, Russia}

\author{A.V. Luchinsky}

\email{Alexey.Luchinsky@ihep.ru}

\begin{abstract}

The module for the generation of $B_c$ meson decays into $J/\psi  + n\pi$ and $B_s^{(*)}  + n\pi$  ($n\le 4$)  is implemented into EvtGen program package. The decay amplitudes are calculated 
in the frame work of factorization model. Within this approach the decay can be represented as  $B_c$ decay into  $J/\psi (B_s)+W^*$ followed by the virtual $W^*$-boson decay into the final set of $\pi$-mesons. The described calculation technique allows to adopt the parameters of $W^*\to n\pi$ transition  from the analysis of $\tau$ decay into $\nu_{\tau}+n\pi$. Comparison with available theoretical predictions is performed.

\end{abstract}

\maketitle


{\bf PROGRAM SUMMARY}

\begin{small}
\noindent
{\em Manuscript Title: Use of EvtGen package for the analysis of $B_c \to V/P +n\pi$ decays}                                       \\
{\em Authors: A.V. Berezhnoy, A.K. Likhoded, A.V. Luchinsky}                                                \\
{\em Program Title: BC\_NPI}                                          \\
{\em Programming language: C++}                                   \\
{\em Operating system: Linux}                                       \\
{\em Keywords:} heavy meson, decay, charmonium  \\
{\em External routines/libraries: EvtGen, ROOT, CLHEP}                                      \\
\end{small}

\section{Introduction}

Recent measurements of $B_{c}$ meson mass and lifetime in CDF \cite{Aaltonen:2007gv}
and D0  \cite{Abazov:2008rb} experiments  are the first steps in the experimental research of heavy quarkonia with open flavor. The obtained experimental results  are in a good agreement with the theoretical predictions for the $B_c$ mass~\cite{Gershtein1995aaa, Eichten:1994gt, Ebert:2002pp}:
$$m_{B_c}^{\rm CDF}=6.2756\pm 0.0029(\textrm{stat.})
\pm 0.0025(\textrm{sys.}) \;\textrm{GeV}, $$
$$m_{B_c}^{\rm D0}=6.3000\pm 0.0014(\textrm{stat.})
\pm 0.0005(\textrm{sys.}) \;\textrm{GeV},
$$
$$ \qquad  m_{B_c}^{\rm theor}=6.25\pm 0.03 \;\textrm{GeV};$$
as well as for the decay time~\cite{Kiselev:1999sc,Kiselev:2000pp,Kiselev:1994ay,Kiselev:1993eb,Kiselev:1992tx}:
$$\tau_{B_c}^{\rm CDF}=0.448^{+0.038}_{-0.036} (\rm stat.)\pm 0.032 (\rm sys.)\;\textrm{ps},$$
$$\tau_{B_c}^{\rm D0}=0.475^{+0.053}_{-0.049} (\rm stat.)\pm 0.018 (\rm sys.)\;\textrm{ps},$$
$$\tau_{B_c}^{\rm theor}=0.48 \pm 0.05 \;\textrm{ps}.$$

The $B_c$ meson was observed only in two decay modes: $B_c\to J/\psi \pi$ and
$B_c\to J/\psi + \mu +\nu_{\mu}$. The investigation of other decay modes will be possible at
at LHC, where about $10^{10}$ events with $B_c$ mesons per year are expected. 
This huge amount of events will allow to obtain the information on the production cross section distributions, on the decay branching fractions, and in some cases, on the distributions of decay products. 

The  $B_c$ systems do not have strong and electromagnetic annihilation decay modes. Due to this reason the exited $B_c$ systems laying below $B+D$ threshold have the decay widths by two order of magnitude smaller  than  the widths for analogous exited states of charmonium and bottonium. All exited $B_c$ states after a set of radiative transitions decay into the lightest pseudoscalar state ($0^-$). The lifetime of this state is comparable with  the lifetimes of  
of $B$ and $D$ mesons and essentially differ from lifetimes of other lightest quarkonia: $\eta_c$   and $\eta_b$.  This is why  $B_c$ meson 
provides a unique possibility to investigate the both  strong and weak interactions.

The main  $B_c$ decay modes, such as  $B_c\to J/\psi+\ell\nu$, $B_c\to J/\psi\pi$ and $B_c\to B_s^{(*)}\rho$  were theoretically studied in details (see, for example \cite{Kiselev:1999sc,Kiselev:2002vz,Wang:2008xt}). For all these processes it is assumed that the factorization approach is valid: the  decay $B_c \to \textrm{heavy  hadron} +W^*$  is followed by the decay of  the virtual $W$-boson. The transition form-factors for the processes $B_c\to \textrm{heavy hadron} +W^*$ can be estimated within QCD sum rules, within quark potential models,  or in the framework of light-cone quark model.

In our recent articles \cite{Likhoded:2009ib, Likhoded:2010jr} we have studied the $B_c$ decay process with several pions in the final state, such as   $B_c\to
J/\psi+n\pi$ and $B_c\to B_s^{(*)}+n\pi$ with $1\le n\le 4$. The factorization approach have been used in these calculations. The characteristics of the virtual $W$-boson decay have been adopted from the experimental data on $\tau\to\nu_\tau+n\pi$ decays for $n\le 3$  and from the experimental data on $\pi$ mesons production in the process $e^+e^-\to 4\pi$.

The modules for calculation of these decay amplitude is implemented into EvtGen program package~\cite{EvtGen}.
Therefore the detailed simulation of these decays in the kinematical condition of real experiments is possible now. In this paper we describe the installation and using the developed modules.

 This paper is organized as follows. In the next section we present a theoretical background for calculation of $B_c\to
V(p)+n\pi$ decays. In sections \ref{sec:soft} the structure of presented packages is described. The installation instruction and the comparison with the available data are given in section \ref{sec:comparison}. The conclusions are given in the final section.

\section{Theoretical background}

\begin{figure}[!t]
\centering
\resizebox*{1.0\textwidth}{!}{  
\includegraphics{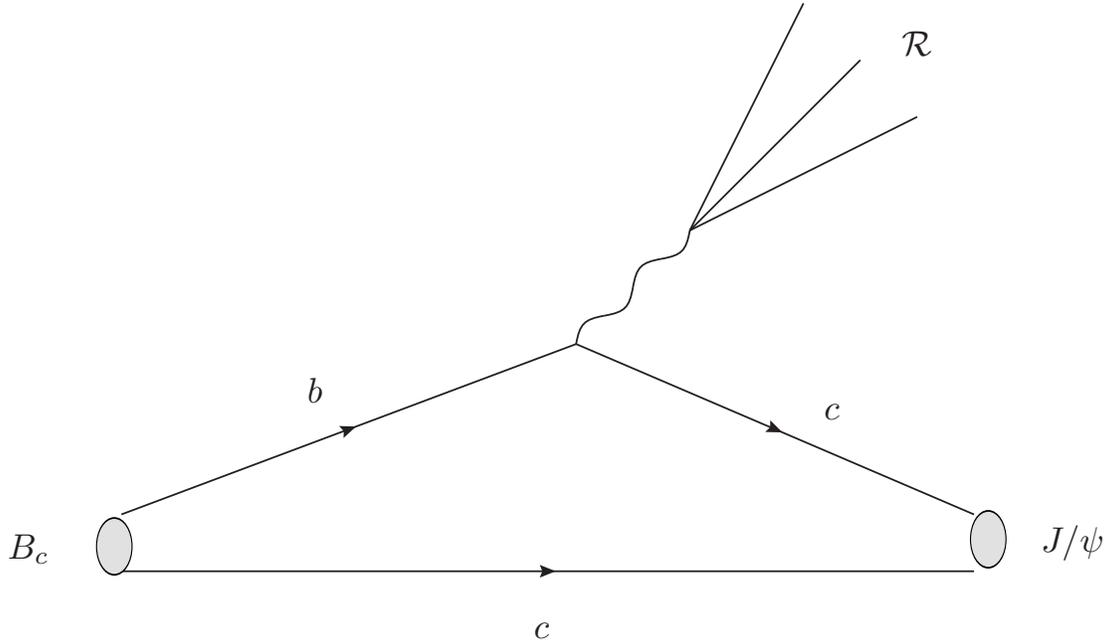}
}
  \caption{Typical diagram for $B_c\to V(P)+n\pi$ decay.}
  \label{fig:diag}
\end{figure}

To calculate the decay amplitudes we  assume that the discussed processes can be 
represented as  the  decay $B_c\to \textrm{heavy  hadron} +W^*$  followed by the decay of  the virtual $W$-boson. An additional assumption should be made  that the only one of the constituent quarks weakly decays, meanwhile the other quark remains the same (a typical diagram is presented in Fig.~\ref{fig:diag}).

Within this approach the amplitude of the process  can be written in the form
\begin{equation}
  \mathcal{A}[B_c \to  J/\psi + n\pi] = \frac{G_F V_{cb}}{\sqrt{2}} a_1(\mu_R) \mathcal{H}^{J/\psi}_\mu \epsilon_W^\mu,
  \label{eq:amp_Jpsi}
\end{equation}
where $\epsilon_W$ is the polarization vector of $W^*$;

$\mathcal{H}^{J/\psi}$ is the $B_c\to J/\psi +
W^*$ transition vertex:
\begin{equation}
\mathcal{H}_{\mu} = \left\langle J/\psi\left|\bar{c}\gamma_{\mu}\left(1-\gamma_{5}\right)b\right|B_{c}\right\rangle =\mathcal{V}_{\mu}-\mathcal{A}_{\mu}.
\end{equation}

Vector and axial currents are equal to
\begin{equation}
\mathcal{V}_{\mu}  =  \left\langle J/\psi\left|\bar{c}\gamma_{\mu}b\right|B_{c}\right\rangle =
i\epsilon^{\mu\nu\alpha\beta}\epsilon_{\nu}^{\psi} p_{\alpha}q_{\beta}F_{V}\left(q^{2}\right),
\end{equation}

\begin{multline}
\mathcal{A}_{\mu}  = \left\langle J/\psi\left|\bar{c}\gamma_{\mu}\gamma_{5}b\right|B_{c}\right\rangle = \\
\epsilon_{\mu}^{\psi}F_{0}^{A}\left(q^{2}\right)+p_{\mu}\left(\epsilon^{\psi}p_{B_c}\right)F_{+}^{A}\left(q^{2}\right)+q_{\mu}\left(\epsilon^{\psi}p_{B_c}\right)F_{-}^{A}\left(q^{2}\right),
\end{multline}
where $p_{B_c}$ and $p_{J/\psi}$ are the momenta of $B_{c}$- and $J/\psi$-mesons;

$q= p_{B_c}-p_{J/\psi}$ is the momentum of virtual $W$-boson;

 $p= p_{B_c}+p_{J/\psi}$;

$\epsilon_\mu^{J/\psi}$ is the polarization vector of $J/\psi$ meson;

and $F_{V}(q^{2})$, 
 $F^A_{0}(q^2)$, $F^A_{+}(q^2)$, $F^A_{-}(q^2)$ and $F_V(q^2)$ are form-factors of $B_c\to J/\psi+W^*$ decays.

In the tree approximation the parameter $a_1(\mu_R)$  is equal to unity. Higher-order corrections lead to dependence of this factor on
renormalization scale $\mu_R$ \cite{Buchalla:1995vs}. Numerical values for $a_1(\mu_R)$ at different scale are  calculated in \cite{Kiselev:2002vz} .
For the process   $B_c\to J/\psi+n\pi$  the value of $\mu_R$ has been chosen to be equal to the  mass value of the decayed $b$-quark:
$$a_1(m_b) = 1.4$$

The form-factors were calculated within different nonperturbative approaches: QCD sum rules \cite{Kiselev:2002vz}, potential quark models \cite{Kiselev:1993eb}, and
Light-Front quark models \cite{Anisimov:1998uk, Huang:2008zg}. In our article we use exponential parametrization of these form-factors:
\begin{eqnarray}
  F_i(q^2) &=& F_i(0) \exp\left\{ c_1 q^2 + c_2 q^4\right\}.
  \label{eq:Fi}
\end{eqnarray}

The reason for this choice is that in the framework of potential quark models the form-factors of $B_c\to J/\psi$ transitions are expressed
through the interception integral of initial and final meson wave functions. For heavy quarkonia these wave functions are usually
parametrized in exponential form, so presented above expression for form-factors is natural. However, it can be shown,  that the numerical
results do not depend strongly on the parametrization form. In the Tab.~\ref{tab:ff} we give numerical values of form-factor parameters
$F_i(0)$, $c_{1,2}$ for different form-factor sets.

\begin{table}
\begin{center}
\begin{tabular}{|c|c|c|c|}
\hline 
\multicolumn{4}{|c|}{$B_c \to J/\psi +n \pi$} \\ \hline
 &$F_i(0)$ & $c_1$ & $c_2$  \\ \hline
 $A_0$ & 5.9 & 0.049 & 0.0015 \\ \hline
 $A_+$ &-0.074 & 0.049 & 0.0015  \\ \hline
 $V$ &	0.11 & 0.049 & 0.0015 \\ \hline  \hline
\multicolumn{4}{|c|}{$B_c \to B_s^* +n \pi$} \\ \hline
 &$F_i(0)$ & $c_1$ & $c_2$  \\ \hline
 $A_0$ & 8.1 & 0.30 & 0.069 \\ \hline
 $A_+$ &0.15 & 0.30 & 0.069  \\ \hline
 $V$ &	1.08 & 0.30 & 0.069 \\ \hline \hline
\multicolumn{4}{|c|}{$B_c \to B_s +n \pi$} \\ \hline
 &$f_i(0)$ & $c_1$ & $c_2$  \\ \hline
 $f_+$ & 1.3  & 0.30  & 0.069   \\ \hline
\end{tabular}
\end{center}
\caption{Form-factor parameters for different SR form-factor sets.}
\label{tab:ff}
\end{table}

The decay  $B_c\to B_s^*+n\pi$ is described by analogy with the decay $B_c\to J/\psi+n\pi$. The same formula for the amplitude is used with the other parameter set (see Tab.~\ref{tab:ff}).

The amplitude for the decay  $B_c\to B_s+n\pi$   can be written in more simple form
\begin{eqnarray}
  \mathcal{A}[B_c \to  B_s + n\pi] &=& \frac{G_F V_{cs}}{\sqrt{2}} a_1(\mu_R) \mathcal{H}^{B_s}_\mu \epsilon_W^\mu,
  \label{eq:amp_Bs}
\end{eqnarray}
where
\begin{equation}
  \mathcal{H}^{B_s}_\mu = \left\langle B_s\left|\bar{c}\left(1-\gamma_{5}\right)b\right|B_{c}\right\rangle
 = f_+(q^2) p_\mu + f_-(q^2) q_\mu.
\end{equation}

The detailed information about $W^*$ decay is not needed  to obtain the integrated branching fractions, as well as the branching fraction distributions on $q^2$. Let us consider the decay 
$B_c\to J/\psi W^* \to J/\psi n \pi$ as an example:

\begin{equation}
d\Gamma\left(B_c\to J/\psi\mathcal{R}\right)  = \\
 \frac{1}{2M}\frac{G_F^{2}V_{cb}^2}{2}a_1^2\mathcal{H}^{\mu}\mathcal{H}^{*\nu}
{\epsilon_{\mu}^W}^*{\epsilon_{\nu}^W}d\Phi\left(B_c\to J/\psi n \pi\right),
\end{equation}

where Lorentz-invariant phase space is defined according to

\begin{equation}
d\Phi\left(Q\to p_{1}\dots p_{n}\right) =  (2\pi)^{4}\delta^{4}\left(Q-\sum p_{i}\right)\prod\frac{d^{3}p_{i}}{2E_{i}(2\pi)^{3}}.
\end{equation}

Using the following recurrent expression for the phase space
\begin{equation}
d\Phi\left(B_{c}\to J/\psi W^* \to J/\psi n \pi \right)  =  \frac{dq^{2}}{2\pi}d\Phi\left(B_{c}\to J/\psi W^{*}\right)d\Phi\left(W^{*}\to n \pi \right)
\end{equation}
 one can perform the integration over phase space of the final state $n \pi$:
\begin{equation}
\frac{1}{2\pi}\int d\Phi\left(W^{*}\to n \pi \right)\epsilon_{\mu}^{W}{\epsilon_{\nu}^W}^*  =  \left(q_{\mu}q_{\nu}-q^{2}g_{\mu\nu}\right)\rho_{T}\left(q^{2}\right)+q_{\mu}q_{\nu}\rho_{L}\left(q^{2}\right),
\end{equation}
where spectral functions $\rho_{T,L}\left(q^{2}\right)$
are universal and can be determined from theoretical and experimental
analysis of some other processes, for example $\tau\to\nu_{\tau} n \pi$
decay or electron-positron annihilation $e^{+}e^{-}\to n \pi$. (See
\cite{Likhoded:2009ib,Likhoded:2010jr,Schael:2005am} for details.)
It is worth to mention, that due to the vector current conservation and the partial axial current conservation
spectral function $\rho_{L}$ for $n\geq 2$ is negligible in almost
whole kinematical region, so  it can be neglected in the estimations for $n\geq 2$.
 For the purposes of our article, however, a more detailed description is
required for the multipion final state.

\begin{figure}[!t]
\centering
\resizebox*{1.0\textwidth}{!}{  
\includegraphics{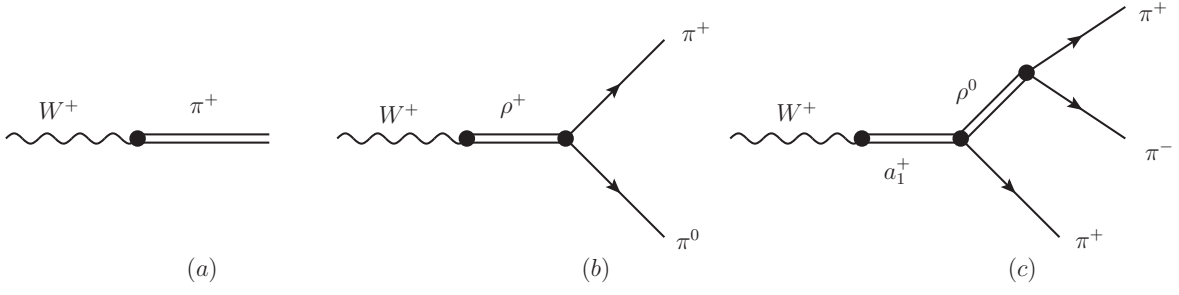}
}
  \caption{Typical diagrams for $W^*\to n\pi$ decay in resonance approximation.}
  \label{fig:digPi}
\end{figure}

In the framework of resonance model the decays of $W^*$-boson can be described in terms of virtual $\rho$- and $a_1$-mesons exchange
(see typical diagrams presented in Fig.~\ref{fig:digPi}).

If there is one $\pi$-meson in the final state (see Fig.~\ref{fig:digPi}a), the vertex of $W^*\to\pi$ transition can be written in the form
\begin{equation}
  \langle \pi^+ |(\bar d u)_{V-A} | W\rangle = f_\pi k_\mu,
\end{equation}
where $k$ is the momentum of $\pi$-meson and  $f_\pi\approx 140$ MeV is its coupling constant. In accordance with this interaction vertex the effective polarization vector $\epsilon^W_\mu$ in this case has the form
\begin{equation}
  \epsilon^W_\mu = f_\pi k_\mu/m_\pi^2.
\end{equation}
Note, that this vertex violates the axial current.

 The decays of virtual $W$-boson into multipion final states  are described within resonance model in terms of virtual $\rho$- and $a_1$-mesons exchange (see typical diagrams presented in Fig.~\ref{fig:digPi}b,c).

The $W^*\to\pi^+\pi^0$ decay is saturated mainly by contributions of virtual $\rho$-
and $\rho'$-mesons (see Fig.~\ref{fig:digPi}b). The corresponding effective polarization vector can be written as
\begin{equation}
  \epsilon^{2\pi}_\mu = F_\rho(q^2) (k_1-k_2)_\mu,
  \label{eq:eps2pi}
\end{equation}
where $k_{1,2}$ are $\pi$-mesons momenta and $F_\rho(q^2)$ is the $\rho$-meson form-factor (see  \cite{Kuhn:1990ad}). The difference in $\pi^0$- and $\pi^+$-meson masses is neglected, thus the virtual $W$ boson in this decay has a transverse polarization. It should be noted that the width of the $\rho$ meson must be taken into account.

The $W^*\to 3\pi$-transition is saturated mainly by $W^*\to a_1\to\rho\pi\to3\pi$
decay chain. Following \cite{Kuhn:1990ad} one can write the effective polarization vertex in this case as
\begin{equation}
  \epsilon^{3\pi}_\mu = -i \frac{2\sqrt{2}}{3f_\pi}F_a(q^2)\left\{
    B_\rho(s_2) V_{1\mu} + B_\rho(s_1) V_{2\mu}
  \right\},
  \label{eq:eps3pi}
\end{equation}
where
\begin{equation}
  V_{1,2\mu} = k_{1,2\mu}-k_{3\mu}-q_\mu \frac{q(k_{1,2}-k_3)}{q^2}
\label{eq:V3pi}
\end{equation}
and
\begin{equation}
  s_{1,2} = (q-k_{1,2})^2.
\end{equation}
Parametrization of $B_\rho(s)$ function is presented in \cite{Kuhn:1990ad}. It can be clearly seen, that if one neglects the difference between charged and neutral $\pi$-meson masses the above expression in transverse and the axial current is conserved.  In EvtGen package this transition was realized already in TAUHADNU model.

The simulation of $W^*\to 4\pi$-transition will be possible within the next version of our package.

\section{Overview of the software structure\label{sec:soft}}

For the Monte-Carlo simulation of the discussed decays the generator \rmbf{EvtGen} has been chosen. This  generator is widely used in high energy physics. For example, \rmbf{EvtGen} is a part of the LHCb software environment  \rmbf{GAUSS}.

The characteristic property of \rmbf{EvtGen} is that, in contrast to many other generators (Pythia, for example), the polarizations of the initial and final particles are taken into account in modeling. This is possible because \rmbf{EvtGen} deals with the amplitude of the process under consideration, while other generators work with averaged over helicities squared matrix elements. As a result, one can easily study different spin asymmetries and other similar quantities.

\begin{figure}[!t]
\centering
\resizebox*{1.0\textwidth}{!}{  
\includegraphics{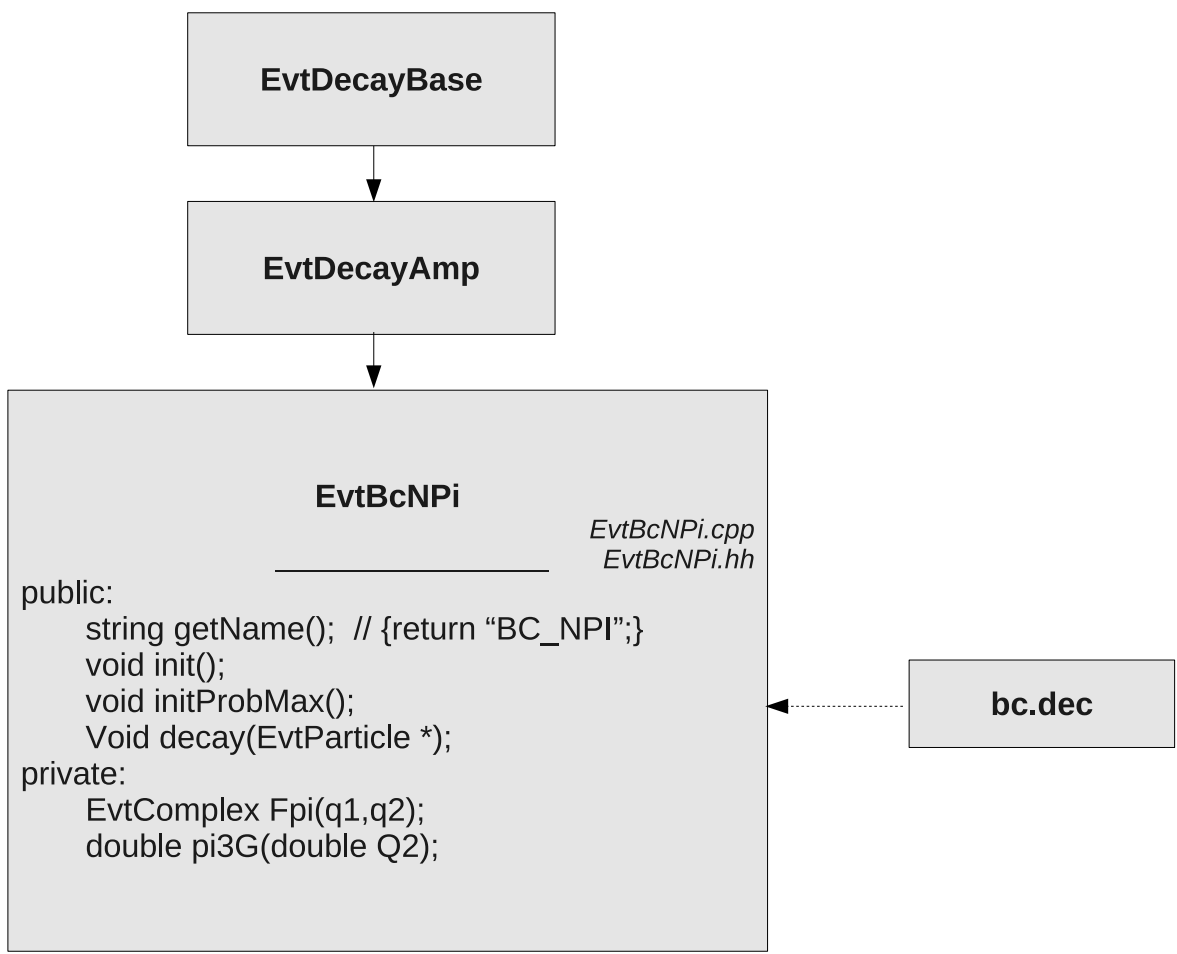}
}
  \caption{The program structure.}
  \label{fig:class}
\end{figure}

\rmbf{EvtGen}-generator is written completely on C++. In order to add a new decay to it one should just create a new class, that describes this decay. The base class is \rmbf{EvtDecayAmp}, where prototypes of all necessary for the model functions are given. The most significant of these functions are \rmbf{init()}, \rmbf{initProbMax()} and \rmbf{decay()} (see Fig.~\ref{fig:class}).

The method \rmbf{init()} performs the initialization of the decay model and reads its parameters. The necessary parameters are stored in the so-called \rmbf{.dec}-file (the exact name of this file can be determined by the user). The detailed description of the .dec-file format can be found in \rmbf{EvtGen} documentation~\cite{EvtGen}. Briefly, the structure of this file is
\begin{flushleft}
\rmbf{
BeginDecay $\left<decParticle\right>$\\
\dots\\
$\left<Br\right>$ $\left<out_1\right>$ \dots $\left<out_n\right>$ $\left<modelName\right>$ $\left<par_1\right>$ \dots $\left<par_m\right>$;\\
\dots\\
EndDecay
}
\end{flushleft}
where
\begin{flushleft}
 $\qquad\left<decName\right>$ is the name of the initial particle, \\
 $\qquad\left<Br\right>$ is the branching fraction of the decay,\\
 $\qquad\left<out_1\right>$ \dots $\left<out_n\right>$ are the names of the final particles,\\
$\qquad\left<modelName\right>$ is the name of the model that describes the specific decay,\\
$\qquad\left<par_1\right>$ \dots $\left<par_m\right>$ are the parameters of this model. The number of these parameters and their values depend on used model and specific decay.
\end{flushleft}
Modifying this .dec-file one can easily switch off the processes that are not currently needed and tune the processes that are used in the program.

When Monte-Carlo generator is used, one needs to determine the maximal probability of the process. If this value is not set correctly, the efficiency of the program can be reduced, In EvtGen package maximum probability is determined by \rmbf{initProbMax} method. If it is not set, the program would try to determine it automatically (simply generating 500 decays and taking the maximum value), and in the subsequent work any overflow incident would be reported.

The main function of any model is \rmbf{decay()} method, where all physics of the process is coded. For a given decay kinematics this method calculates the amplitude of the process for all polarizations of initial and final particles.

In order to include described in the previous section decays $B_c \to V(P) + n \pi$ the model \rmbf{BC\_NPI} was written. This model is realized in class \rmbf{EvtBcNPi} and describes  $B_c$-meson decays both into vector and pseudoscalar mesons. The number of final $\pi$-mesons are limited to $n\le 3$. The base class for both models is EvtDecayAmp (see fig.\ref{fig:class}), so polarizations of initial and final particles are taken into account accurately. The decay()-methods are written according to relations (\ref{eq:eps2pi}), (\ref{eq:eps3pi}), (\ref{eq:V3pi}) from the previous section, and parameters of the models are read from determined by the user .dec-file.

If we are describing $B_c\to V+n\pi$ decay, the corresponding record in the decay file should have the form
\begin{flushleft}
$\left<Br\right>$ $\left<vec\right>$ $\left<pi_1\right>$ \dots $\left<pi_n\right>$ BC\_NPI\\
\qquad $\left<maxProb\right>$\qquad \# maxProb \\
\qquad $\left<F^A_0(0)\right>$ $\left<c_1^{A0}\right>$ $\left<c_2^{A0}\right>$ \qquad \# FA0 \\
\qquad $\left<F^A_+(0)\right>$ $\left<c_1^{A+}\right>$ $\left<c_2^{A+}\right>$ \qquad \# FA+ \\
\qquad $\left<F_V(0)\right>$ $\left<c_1^{V}\right>$ $\left<c_2^{V}\right>$ \qquad \# FV \\
\end{flushleft}
where $\left<Br\right>$ is the branching fraction of the decay, $\left<vec\right>$ is the name of final vector particle, $\left<pi_i\right>$ are the names of final $\pi$-mesons (pi+, pi- and pi0 for positively charged, negatively charge and neutral $\pi$-mesons respectively), and $\left<F_i(0)\right>$, $\left<c_1^i\right>$ and $\left<c_2^i\right>$ are parameters of decay form factor $F_i(q^2)$ (see eq.(\ref{eq:Fi}) ). Note that lines in the decay file can be splitted freely and all symbols after \# character are treated as comments. The order of the parameters, however, is important. In the case of the pseudoscalar meson in the final state the corresponding record in the decay file should have the form
\begin{flushleft}
$\left<Br\right>$ $\left<vec\right>$ $\left<pi_1\right>$ \dots $\left<pi_n\right>$ BC\_NPI\\
\qquad $\left<maxProb\right>$\qquad \# maxProb \\
\qquad $\left<F^+(0)\right>$ $\left<c_1^{F+}\right>$ $\left<c_2^{F+}\right>$ \qquad \# F+ \\
\end{flushleft}
Note that in comparison with the previous example only one form-factor is left.

To clarify this point let us consider the .dec file where only  $B_c\to J/\psi \pi^+ \pi^0$ and $B_c \to B_s \pi^+$ are allowed. If SR form-factors set is used, this decay file should have the form
\begin{verbatim}
  Decay Bc+
# Bc -> J/psi pi+ pi0, SR form-factors set
      0.0017  J/psi pi+ pi0     BC_NPI   
                330.                  # maxProp
                5.9 0.049 0.0015      # FA0
                -0.074 0.049 0.0015   # FAp
                0.11 0.049 0.0015;     # FV

# Bc -> Bs pi+ , SR form-factors set
      0.18     B_s0  pi+        BC_NPI
                250                   #maxProb
                1.3      0.30   0.069;			# Fp
  Enddecay
\end{verbatim}

\section{Installation instructions and comparison with available data\label{sec:comparison}}
 
The EvtGen package requires \rmbf{CERNLIB}, \rmbf{ROOT} and \rmbf{CLHEP} libraries to be properly installed. The location of the corresponding files can be altered by modifying the \rmbf{Makefile}. Unfortunately the original package EvtGen is not supported now. Each experiment involved in the heavy quark research use its own branch of EvtGen for the decay generation. This why our module should be adapted to the concrete EvtGen branch.

Here we present the variant of BC\_NPI module which works with the last original version version of  EvtGen~\cite{EvtGen}. This module version can be send on request by e-mail.

The BC\_NPI module is distributed as a source files \rmbf{EvtBcNpi.cc}, \rmbf{EvtBcNpi.hh}, sample decay file \rmbf{bc.dec}, a simple test program \rmbf{test.cc}, and a \rmbf{Makefile} needed to include the considered model into EvtGen package. For installation of the model one should execute commands
\begin{verbatim}
  $ make lib
  $ sudo make install
  $ make clean
\end{verbatim}

\begin{figure}
\begin{center}
\resizebox*{1.0\textwidth}{!}{\includegraphics{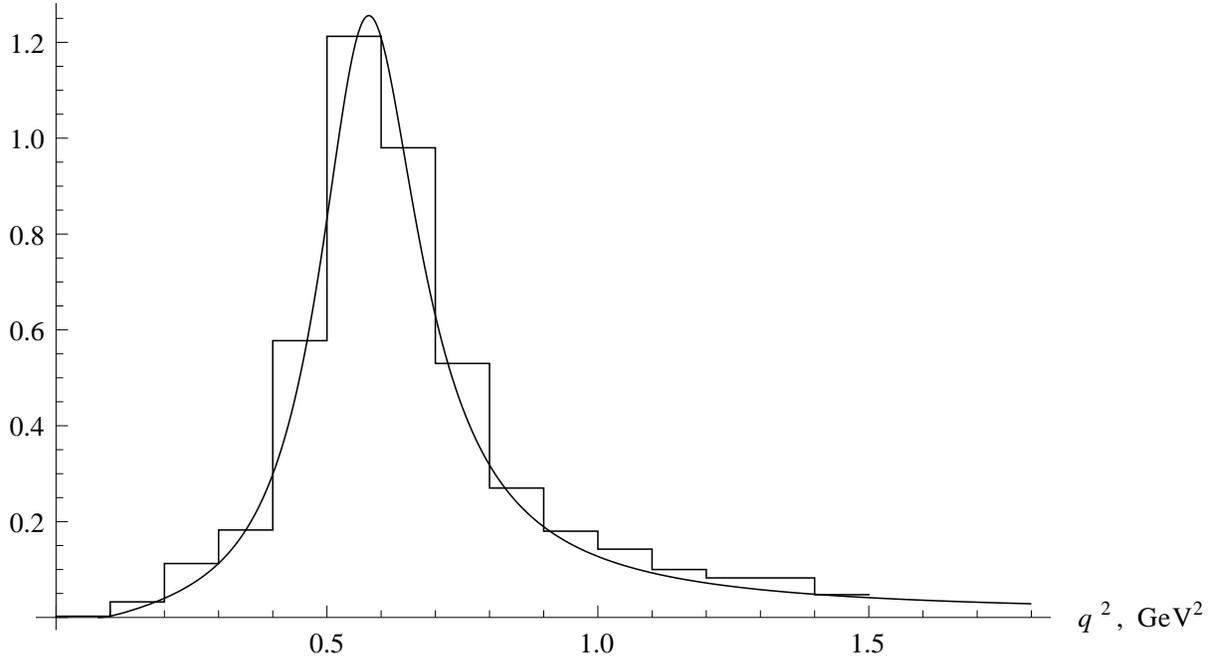}} 
\end{center}
\caption{The distribution over the squared invariant mass $q^2=m_{\pi\pi}^2$ in $B_c\to J/\psi+2\pi$ decay generated within BC\_NPI-model (histogram) in comparison with the predictions of  \cite{Likhoded:2009ib}(curve).\label{fig:comp}}
\end{figure}

Executing commands
\begin{verbatim}
 $ make test
 $ ./test.exe
\end{verbatim}
one can run simple test program. This program generates 10 000 decays $B_c \to J/\psi \pi^0 \pi^+$ and saves a distribution over the invariant mass of the $\pi^0\pi^+$-pair into file \rmbf{hist.root}. The resulting historgam can be viewed with usual ROOT commands
\begin{verbatim}
 $ root hist.root
   hist->Draw()
\end{verbatim}
and should look like the distribution shown in fig.\ref{fig:comp}. For comparison in this figure the results presented in \cite{Likhoded:2009ib} are also shown.
To remove the model from EvtGen package one can execute the command
\begin{verbatim}
 $ sudo make uninstall
\end{verbatim}

In addition, one should modify the decay-file and set the form-factor parameters. The format of this file was discussed in the previous section and parameters of $B_c \to J/\psi$ and $B_c \to B_s^{(*)}$ are presented in table \ref{tab:ff}.

In Fig.~\ref{fig:comp} we show the distribution over the invariant mass of light meson system in $B_c\to J/\psi + 2\pi$ decay is given in comparison with the prediction of work \cite{Likhoded:2009ib}. One can easily see, that the results of Monte-Carlo are in good agreement with available data. For illustration we also show the same distribution with additional cut $p_T^2 > 5\,\mathrm{GeV}^2$, where $p_T$ is the transverse momentum of $J/\psi$-meson in the rest frame of initial $B_c$. Such distributions, obviously, cannot be expressed in terms of spectral functions only.

\section{Conclusion}


$B_c$-mesons, that is particles build from $b$- and $c$-quarks are extremely interesting both from theoretical and experimental points of view. Since the masses of the constituent quarks are large in comparison with $\Lambda_\mathrm{QCD}$, these particles give the opportunity to study QCD both in perturbative and nonperturbative regimes. Decays of $B_c$-mesons, on the other hand, can be caused only by weak interaction, so these particles give also information about electroweak sector of the Standard Model. Theoretically $B_c$-mesons were studied rather thoroughly already, predictions for their  masses, lifetimes and branching fractions of some, mainly two-body, decay modes are  available. In our recent articles \cite{Likhoded:2009ib, Likhoded:2010jr} we study also exclusive decays into a  heavy quarkonia and a system of light mesons. In these articles the spectral function formalism is applied, when the phase space of this system is integrated out. It is clear, that main information about the dynamics of the processes  is lost in this integration, so only branching fractions of the decays and distributions over the invariant mass of $\pi$-mesons can be obtained in the framework of this method. Additional information is required in the conditions of real experiments.

Experimentally rather little is known currently about $B_c$ mesons. Only the mass of the ground state, its lifetime and branching fraction of two decay modes ($B_c\to J/\psi \ell \nu$ and $B_c \to J/\psi\pi$) were measured. According to theoretical estimates, a large yeld of $B_c$-mesons is expected at LHC, so observation of other decay modes is possible. It is clear that, for the analysis of experimental data, it is necessary to have reliable theoretical predictions.

In this paper we consider the exclusive decays $ B_c \to V(P) W^* \to V (P) + n \pi$, where $ 1 \le n \le 3 $. Unlike our previous works \cite{Likhoded:2009ib, Likhoded:2010jr}, we do not use the formalism of spectral functions, and describe in detail the process of transition of virtual $W$-boson in a system of light mesons. This is realized with the help of the model for EvtGen generation package, that allows one to model these decays with different parametrizations of $B_c \to V(P)$ transition form-factors. This model is discussed in details in the present paper.

This research is partially supported by Russian Foundation for Basic Research (grant 10-02-00061a). The work of A.~V.~Luchinsky was also supported by non-commercial foundation ''Dynasty'' and the grant of the president of Russian Federation for young scientists with PhD degree (grant MK-406.2010.2).

\end{document}